\documentclass[reprint,amsmath,amssymb,nofootinbib]{revtex4-2}
\usepackage{xtab,afterpage}
\usepackage{graphicx}
\usepackage{dcolumn}
\usepackage{bm}
\usepackage{natbib}
\usepackage{url}
\usepackage{textcase}
\usepackage{amsmath,amssymb,amsfonts,mathrsfs}
\usepackage{mathtools}
\usepackage{braket}
\usepackage{enumitem} 
\usepackage{makecell}
\usepackage{cases}
\usepackage{babel}
\usepackage{amsthm}

\newcommand{\f}[2]{\frac{#1}{#2}}
\newcommand{\ko}[1]{\left( #1 \right)}
\newcommand{\kko}[1]{\left[ #1 \right]}

\newcommand{\abs}[1]{\left| #1 \right|}
\newcommand{\q}[1]{`#1'}
\newcommand{\dd}{\mathop{}\!d}
\newtheorem{theorem}{Theorem}

\DeclarePairedDelimiter\ceil{\lceil}{\rceil}
\DeclarePairedDelimiter\floor{\lfloor}{\rfloor}
\def\no{\nonumber}
\def\P{\bm{P}}
\def\vK{\bm{k}}
\def\vD{\bm{D}}
\def\vI{\bm{I}}
\def\vA{\bm{A}}
\def\vB{\bm{B}}

\begin{document}

\title{Emergent family of Tsallis entropies from the \textit{q}-deformed combinatorics}

\author{Keisuke Okamura}
\altaffiliation{Embassy of Japan in the United States of America, 2520 Massachusetts Avenue NW, Washington, DC 20008, USA}
\email{okamura@alumni.lse.ac.uk}
\date{September 28, 2024}

\begin{abstract}
We revisit the derivation of a formula for the $q$-generalised multinomial coefficient rooted in the $q$-deformed algebra, a foundational framework in the study of nonextensive statistics.
Previous approximate expressions in the literature diverge as $q$ approaches 2 (or 0, depending on convention).
In contrast, our derived formula provides an exact, smooth function for all real values of $q$, expressed as an infinite series expansion involving Tsallis entropies with sequential entropic indices, coupled with Bernoulli numbers.
This formulation is achieved through the analytic continuation of the Riemann zeta function, stemming from the $q$-deformed factorials.
Our formula thus offers a distinctive characterisation of Tsallis entropy within the $q$-deformed combinatorics.
Throughout this exploration, we also highlight a symmetry within the $q$-deformed theory that links different values of the entropic parameter.
Furthermore, we discuss extending our results to encompass more general, affinity-sensitive cases, building on the previously established framework of affinity-based extended entropy.
\end{abstract}

\maketitle

\section{Introduction\label{sec:intro}}

As is well known, an asymptotic correspondence exists between Shannon's information entropy \cite{Shannon48} and a multinomial coefficient.
To illustrate this, consider, on the one hand, a discrete probability distribution $\P=\{p_{i}\}_{i=1}^{r}$, where $p_{i}\in [0,1]$ and $\sum_{i=1}^{r}p_{i}=1$.
The Shannon entropy of distribution $\P$ is then defined as\footnote{By convention, $p\ln p=p\ln(1/p)=0$ when $p=0$.}
\begin{equation}\label{def:Shannon_ent}
H_{1}(\P)
\coloneqq -\sum_{i=1}^{r}p_{i}\ln p_{i}\,.
\end{equation}
In physics, an entropy formula of the same mathematical form, Gibbs entropy \cite{Gibbs02}, plays a pivotal role in Boltzmann--Gibbs statistical mechanics, bridging the laws of the microscopic physical world with the macroscopic description of the same physical world by thermodynamics.
The entropy formula (\ref{def:Shannon_ent}), referred to as the Boltzmann--Gibbs--Shannon (BGS) entropy, has found applications in various fields of the natural and social sciences through Jaynes's principle of maximum entropy \cite{Jaynes57}, providing a profound lens through which to understand natural phenomena and their underlying mechanisms.

On the other hand, consider the multinomial coefficient defined as
\begin{equation}\label{def:multinomial}
\binom{n}{\vK}\equiv\binom{n}{k_{1},\,\dots,\,k_{r}}\coloneqq 
\f{n!}{\prod_{i=1}^{r}k_{i}!}\,,
\end{equation}
where $\vK=\{k_{i}\}_{i=1}^{r}$ represents a set of nonnegative integers and $n=\sum_{i=1}^{r}k_{i}$.
This quantity calculates the number of distinct permutations in a multiset of $r$ distinct categories with multiplicities $\vK$.
In a physical scenario, consider $n$ particles that can occupy $r$ distinct states, such as energy levels, where a macrostate is characterised by having $k_i$ particles in the $i$-th state for each $i$. 
The multinomial coefficient (\ref{def:multinomial}) then represents the number of possible microstates corresponding to a given macrostate specified by $\vK$. 
The greater the multinomial coefficient, the higher the probability of that macrostate occurring.

Suppose that $n$ and each $k_{i}\in\vK$ are of the same order and are large enough to be effectively infinite (the \q{asymptotic limit}).
Taking the logarithm of the multinomial coefficient (\ref{def:multinomial}) and using Stirling's formula of the form $\ln(n!)=n\ln n-n+O(\ln n)$ for $n$ and each $k_{i}$, it is straightforward to verify that
\begin{equation}\label{ln_multicoeff}
\ln\binom{n}{\vK}
=nH_{1}(\vK/n)+{O}(\ln n)\,.
\end{equation}
We observe that BGS entropy (\ref{def:Shannon_ent}) emerges directly from the multinomial coefficient (\ref{def:multinomial}) in the leading term proportional to $n$.
This widely acknowledged observation highlights a significant structural connection between the most prevailing entropy measure, BGS entropy, and one of the fundamental combinatorial quantities, the multinomial coefficient.
This naturally prompts the question of whether a similar or more general relationship exists in contexts beyond BGS entropy.

This paper aims to contribute to this ongoing exploration.
Specifically, we explore in detail which functionals of generalised entropies appear on the right-hand side (RHS) when the multinomial coefficient and its logarithm on the left-hand side (LHS) are replaced with their $q$-deformed versions \cite{Tsallis88,Tsallis94,Nivanen03,Borges04}, both of which are described later (see Eqs.~(\ref{def:q-multinomial1},\,\ref{def:q_ln})).
We aim to derive a continuous function $L_{q}$ of $\P$ that is defined over the entire domain of $q\in\mathbb{R}$ and asymptotically represents the $q$-logarithm of the $q$-deformed multinomial coefficient:
\begin{equation}
\ln_{q}\binom{n}{\vK}_{\!q}\sim L_{q}(\P)\qquad (n\sim k_{i}\gg 1)\,.
\end{equation}
Some prior literature has taken important steps in this direction \cite{Suyari06,Oikonomou07}.
However, our understanding requires further refinement and completion.
For instance, in the literature \cite[Eq.~(42)]{Suyari06}, the expressions for the function $L_{q}$ were obtained by dividing the cases as follows:
\begin{subnumcases}{\label{Sy_lnq_multi} 
L_{q}(\P)~\stackrel{\raisebox{1ex}{\text{\footnotesize\cite{Suyari06}}}}{\approx}~}
\,\f{n^{2-q}}{2-q}H_{2-q}(\P) & ~\text{$(q>0,~\,q\neq 2)$}\,, \label{Sy_lnq_multi_a}  \\
\,-\ln\ko{\f{n}{\prod_{i=1}^{r}k_{i}}} & ~\text{$(q=2)$}\,, \label{Sy_lnq_multi_b}
\end{subnumcases}
where $H_{q}(\P)$ denotes the Tsallis entropy function with entropic parameter $q$ (see Eq.~(\ref{def:Tsallis_ent}) for definition). 
The RHS of Eq.~(\ref{Sy_lnq_multi_a}) diverges to $\pm\infty$ as $q\to 2^{\pm 0}$ due to a simple pole at $q=2$, and the function is not continuous at $q=2$.
In addition, while the RHS of Eq.~(\ref{Sy_lnq_multi_a}) yields a negative value for $q>2$, the LHS must remain positive to be consistently associated with the \q{(effective) number of permutations} or the \q{(effective) number of possible microstates} in a physically relevant sense, even when $q\neq 1$.

In contrast, this paper demonstrates that the function $L_{q}$ can be expressed as an analytic, smooth function of $q\in\mathbb{R}$, expandable in the following infinite series form:
\begin{equation}\label{formula}
L_{q}(\P)
=\f{\zeta(q-1)}{q-1}H_{0}(\P)+\sum_{\ell=0}^{\infty}C_{\ell}(q)n^{2-q-\ell}H_{2-q-\ell}(\P)\,,
\end{equation}
where $\zeta(\cdot)$ denotes the Riemann zeta function and $\{C_{\ell}(q)\}_{\ell\in\mathbb{Z}_{\geq 0}}$ are appropriate coefficients involving Bernoulli numbers.
The value at $q=1$ is defined by the limit as $q$ approaches 1.
As we will see, this function's analyticity is derived from the analytic continuation of the Riemann zeta function.
The formula of the form (\ref{formula}) is our main result in this paper, which holds significant potential for advancing our understanding of generalised entropies and their relationship with combinatorial quantities.

The rest of this paper is organised as follows. 
Section \ref{sec:preliminaries} summarises the definition and basic properties of Tsallis entropy.
It also outlines key definitions of the $q$-algebra, including $q$-multiplication and $q$-division, which are foundational for introducing $q$-factorial and $q$-multinomial coefficients.
In addition, we discuss a special symmetry inherent in the $q$-algebra that connects different values of $q$ within the same theoretical framework.
In Section \ref{sec:emergent}, we derive the exact formula (\ref{formula}) and explore the behaviours of the function at characteristic values and limits.
Subsequently, in Section \ref{sec:affinity}, we discuss the potential extension of our results to encompass more general, affinity-sensitive cases based on the previously established framework of distance- or affinity-based extended entropy \cite{Rao82,Leinster12,Okamura20}.
Finally, Section \ref{sec:conclusion} presents a summary and concluding remarks.

\section{Preliminaries\label{sec:preliminaries}}

This section prepares for deriving the $q$-logarithm of the $q$-deformed multinomial coefficient, covering the fundamental properties of Tsallis entropy and its underlying $q$-algebraic framework.

\subsection{Tsallis entropy\label{subsec:Tsallis}}

For over half a century, the success of BGS entropy (\ref{def:Shannon_ent}) has driven researchers to extend theoretically consistent frameworks based on information-theoretic or physical motivations \cite{VonNeumann27,Kullback51,Renyi61,Havrda67,Daroczy70,Sharma75,Tsallis88,Kaniadakis01}.
For a recent review, see e.g.~\cite{Tsallis22} and references therein.
One of the most studied examples is the Tsallis formalism \cite{Tsallis88,Curado91}, in which the concept of Tsallis entropy \cite{Havrda67,Daroczy70,Tsallis88} plays a pivotal role. 
This first example of nonadditive entropy generalises the ordinary additive BGS entropy, significantly advancing nonextensive statistical mechanics and beyond \cite{TsallisBibList}.
Mathematically, Tsallis entropy and its associated theory rely on the following definitions of the $q$-deformed logarithmic and exponential functions \cite{Tsallis94} with the entropic parameter $q\in\mathbb{R}\backslash\{1\}$:
\begin{alignat}{3}
\ln_{q}x&\coloneqq
\f{x^{1-q}-1}{1-q}
	&\qquad& (x>0)\,,\label{def:q_ln}\\
\exp_{q}x&\coloneqq 
\big[1+(1-q)x\big]_{+}^{\f{1}{1-q}}
	&\qquad& (x\in\mathbb{R})\,,\label{def:q_exp}
\end{alignat}
where $[\,\cdot\,]_{+}\coloneqq \max\{\,\cdot\,,0\}$.
The $q$-logarithmic function is understood as $\ln_{q}x\coloneqq\int_{1}^{x}\dd s/s^{q}$, and the $q$-exponential function is its inverse function, satisfying $\ln_{q}(\exp_{q}x)=\exp_{q}(\ln_{q}x)=x$ for appropriately defined $x$.
These are generalisations of the standard logarithmic and exponential functions, which in the limit $q\to 1$ yield $\ln x$ and $\exp{x}$, respectively.
Using the $q$-logarithm (\ref{def:q_ln}), Tsallis entropy is defined as
\begin{equation}\label{def:Tsallis_ent}
H_{q}(\P)\coloneqq
-\sum_{i=1}^{r}(p_{i})^{q}\ln_{q}p_{i}
=\f{1}{1-q}\kko{\sum_{i=1}^{r}(p_{i})^{q}-1}\,,
\end{equation}
and its limit as $q$ approaches 1, $\lim_{q\to 1}H_{q}=H_{1}$, gives the BGS entropy (\ref{def:Shannon_ent}).
Additionally, note that $H_{q}=\sum_{i=1}^{r}p_{i}\ln_{q}(1/p_{i})=-\sum_{i=1}^{r}p_{i}\ln_{2-q}p_{i}$, and $H_{0}=\abs{\P}-1=r-1$.

The entropic parameter $q$ characterises the degree of nonadditivity induced by Tsallis entropy.
Consider two probabilistically independent systems with respective probability distributions $\P_{\mathrm{A}}=\{p_{i}^{\mathrm{A}}\}$ and $\P_{\mathrm{B}}=\{p_{j}^{\mathrm{B}}\}$.
The Tsallis entropy of the joint system, with the joint probability distribution $\P_{\mathrm{A}\cup\mathrm{B}}=\{p_{ij}^{\mathrm{A}\cup\mathrm{B}}\coloneqq p_{i}^{\mathrm{A}}p_{j}^{\mathrm{B}}\}$, satisfies
\begin{equation}\label{nonadditive}
H_{q}(\P_{\mathrm{A}\cup\mathrm{B}})=H_{q}(\P_{\mathrm{A}})+H_{q}(\P_{\mathrm{B}})+(1-q)H_{q}(\P_{\mathrm{A}})H_{q}(\P_{\mathrm{B}})\,.
\end{equation}
This contrasts with the additive property of BGS entropy, where $H_{1}(\P_{\mathrm{A}\cup\mathrm{B}})=H_{1}(\P_{\mathrm{A}})+H_{1}(\P_{\mathrm{B}})$. 
With this nontrivial entropic parameter, Tsallis entropy (\ref{def:Tsallis_ent}) can address the limitations of BGS entropy (\ref{def:Shannon_ent}) in studying long-range interactions, long-time memories, or (multi)fractal structures of systems \cite{Tsallis88,Abe01} that cannot be captured by the BGS statistics.

The possible physical interpretation or origin of the entropic parameter $q$ has been discussed extensively in the literature \cite{Plastino94,Wilk00,Almeida01,Olavo01,Adib03,Biro14,Plastino23}.
Note also that Tsallis entropy is concave for $q>0$, flat for $q=0$ (i.e.~$H_{0}=r-1=\text{const.}$) and convex for $q<0$.
While different studies focus on various domains of $q$---for instance, positive or nonnegative real numbers, or some finite intervals---based on their specific motivations or contexts, our current study considers the entire range of real values of $q$, including $q=1$ as a limiting case.

\subsection{The \textit{q}-algebra and \textit{q}-combinatorics\label{subsec:q-algebra}}

The nonadditive relation (\ref{nonadditive}) has motivated researchers to construct generalised algebraic and combinatorial frameworks \cite{Tsallis94,Abe97,Nivanen03,Borges04,Niven09,Borges22} in which the $q$-logarithmic and $q$-exponential functions (\ref{def:q_ln},\,\ref{def:q_exp}) become natural operations. 
This section, along with Appendix \ref{app:q-algebra}, summarises the definitions and basic properties related to $q$-multiplication, $q$-division, $q$-addition, $q$-subtraction, and their interrelationships.
Particular emphasis is placed on the underlying symmetry related to the deformation parameter ($q$) common to these operations.

\paragraph{The ${q}$-multiplication and ${q}$-division.}

We first review the definition and properties of the $q$-multiplication and the $q$-division \cite{Nivanen03,Borges04}, inspired by Tsallis's nonextensive statistical mechanics formalism \cite{Tsallis88,Curado91,Abe01}. 
These operations are defined between two positive numbers $x$ and $y$ with $q\in\mathbb{R}\backslash\{1\}$ as:
\begin{equation}\label{def:q_multidiv}
x\,\Big\{{}^{\displaystyle\otimes_{q}}_{\displaystyle\oslash_{q}}\Big\}\,y
\coloneqq
\big[x^{1-q}\pm y^{1-q}\mp 1\big]_{+}^{\f{1}{1-q}}
	\qquad (x,\,y>0)\,,
\end{equation}
where the upper signs on the RHS correspond to the definition of $q$-multiplication ($\otimes_{q}$), while the lower signs correspond to $q$-division ($\oslash_{q}$).
These operations reduce to ordinary multiplication and division in the limit $q\to 1$.
By definition, $q$-multiplication is commutative ($x\otimes_{q}y=y\otimes_{q}x$) and associative ($x\otimes_{q}(y\otimes_{q}z)=(x\otimes_{q}y)\otimes_{q}z$).
It also follows that
\begin{align}\label{q_prod_n}
&\bigotimes_{i=1}^{n}{\!}_{q}\,x_{i}
=\exp_{q}\sum_{i=1}^{n}\ln_{q}x_{i}
=\kko{\sum_{i=1}^{n}x_{i}^{1-q}-(n-1)}_{+}^{\f{1}{1-q}}\no\\
&\quad\Rightarrow\quad 
\ln_{q}\bigotimes_{i=1}^{n}{\!}_{q}\,x_{i}
=\sum_{i=1}^{n}\ln_{q}x_{i}\,.
\end{align}

\paragraph{The symmetry.}

There is a remarkable symmetry associated with the $q$-multiplication and $q$-division.
For any $\theta\in\mathbb{R}\backslash\{0\}$, the following relation holds \cite{Korbel17}:
\begin{equation}\label{lam_symmetry}
\big(x^{\theta}\odot_{q}y^{\theta}\big)^{\f{1}{\theta}}
=x\odot_{q'}y~\quad
(\odot=\otimes,\,\oslash)\,,
\end{equation}
where the transformed parameter $q'$ is given by
\begin{align}\label{newlambda}
q'=1+\theta(q-1)\eqqcolon f_{-\theta}(q)\,.
\end{align}
The symmetry induced by the transformation (\ref{newlambda}), referred to as the $\theta$-symmetry in this paper, defines a one-dimensional group that satisfies the axioms of closure, associativity, invertibility and identity \cite{Korbel17}.
It underscores the consistent transformative properties of the $q$-algebra, ensuring that $q$-multiplication and $q$-division retain their forms under the transformation.
In other words, the effect of introducing $\theta$ can be absorbed into a new definition of the entropic parameter.\footnote{Note that what is referred to as the \q{$(\mu,\nu,q)$-relation} in \cite[Eq.~(5) or Eq.~(53)]{Suyari08} simply refers to the relation (\ref{newlambda}), where the triplet $(\mu,\nu,q)$ in the literature corresponds to $(q,-\theta,q')$ in Eq.~(\ref{newlambda}).
It corresponds to the $\theta$-symmetry that already manifests itself in the definition of $q$-multiplication/division as a binary operation (and of $q$-addition/subtraction; see Eq.~(\ref{lam_symmetry2}) in Appendix \ref{app:q-algebra}), not something that emerges specifically in the analysis of the $q$-factorial, let alone the $q$-multinomial coefficient.}
Interesting special values of $\theta$ include $\theta=-1$, corresponding to the \q{additive duality} $q\leftrightarrow 2-q=f_{1}(q)$, and $\theta=-\f{1}{q}$, corresponding to the \q{multiplicative duality} $q\leftrightarrow \f{1}{q}=f_{\f{1}{q}}(q)$ \cite{Curado91,Naudts04,Tsallis17}.
Explicitly, these duality relations are expressed as ($\odot=\otimes,\,\oslash$):
\begin{align}\label{lam_symmetry_spec}
\big(x^{-1}\odot_{q}y^{-1}\big)^{-1}
&=x\odot_{2-q}y\,,\\
\big(x^{-\f{1}{q}}\odot_{q}y^{-\f{1}{q}}\big)^{-q}
&=x\odot_{\f{1}{q}}y\,,
\end{align}
respectively.
Further, it is also straightforward to see that
\begin{equation}\label{lambda_trick}
\f{1}{\theta}\ln_{q}\big(x^{\theta}\big)=\ln_{q'}x
\quad\text{and}\quad
\big[\exp_{q}(\theta x)\big]^{\f{1}{\theta}}=\exp_{q'}x
\end{equation}
for appropriately defined $x$, which directly follow from the definition of the $q$-deformed logarithmic and exponential functions (\ref{def:q_ln},\,\ref{def:q_exp}).

Some possible physical interpretations of this transformation are discussed in \cite{Beck02,Korbel17}.
Note that, as can be seen, even if one starts with positive $q$, the new entropic parameter $q'$ after the transformation (\ref{newlambda}) can be either positive or negative, depending on the domain of $\theta$, with the only exception being the case with the fixed point $q=1$.
We also point out that this transformation function naturally appears in the definition of the $q$-addition and $q$-subtraction as well (see Eq.~(\ref{def2:q_addsubt})).

\paragraph{The ${q}$-factorial.}

Using the $q$-multiplication (\ref{def:q_multidiv}), the $q$-factorial, denoted as $n!_{q}$, is defined by \cite{Suyari06}
\begin{align}\label{def:!q}
n!_{q}\coloneqq 
\bigotimes_{\ell=1}^{n}{\!}_{q}\,\ell
&=\exp_{q}\sum_{\ell=1}^{n}\ln_{q}\ell\no\\
&=\big[h_{n}(q-1)-(n-1)\big]_{+}^{\f{1}{1-q}}\,,
\end{align}
which reproduces the ordinary factorial $n!$ in the limit $q\to 1$.
Here,
\begin{equation}\label{def:harmonic}
h_{n}(s)\coloneqq 
\sum_{\ell=1}^{n}\f{1}{\ell^{s}}
=1+\f{1}{2^{s}}+\f{1}{3^{s}}+\dots+\f{1}{n^{s}}\,,
\end{equation}
denotes the generalised harmonic number defined for $s\in\mathbb{R}$, with the case of $s=1$ corresponding to the ordinary harmonic number.
In the limit as $n\to\infty$, Eq.~(\ref{def:harmonic}) defines the Riemann zeta function for $s>1$, a fact we will utilise later.

It is noteworthy that one can extend the definition of the $q$-factorial using the earlier discussed $\theta$-symmetry:
\begin{equation}\label{def:!q_gen}
n!_{q;\theta}\coloneqq 
\kko{\bigotimes_{\ell=1}^{n}{\!}_{q}\,\ell^{\theta}}^{\f{1}{\theta}}
~\quad (\theta\in\mathbb{R}\backslash\{0\})\,.
\end{equation}
Using the relations (\ref{q_prod_n},\,\ref{lambda_trick}), one can verify that $n!_{q;\theta}=n!_{q'}$ holds under the transformation (\ref{newlambda}).
Explicitly,
\begin{align}
\kko{\bigotimes_{\ell=1}^{n}{\!}_{q}\,\ell^{\theta}}^{\f{1}{\theta}}
&\stackrel{\raisebox{1ex}{~\text{\footnotesize (\ref{q_prod_n})}~}}{=}
\kko{\exp_{q}\sum_{\ell=1}^{n}\ln_{q}\ell^{\theta}}^{\f{1}{\theta}}\no\\
&\stackrel{\raisebox{1ex}{~\text{\footnotesize (\ref{lambda_trick})}~}}{=}
\exp_{q'}\kko{\f{1}{\theta}\sum_{\ell=1}^{n}\ln_{q}\ell^{\theta}}\no\\
&\stackrel{\raisebox{1ex}{~\text{\footnotesize (\ref{lambda_trick})}~}}{=}
\exp_{q'}\sum_{\ell=1}^{n}\ln_{q'}\ell
\stackrel{\raisebox{1ex}{~\text{\footnotesize (\ref{q_prod_n})}~}}{=}
\bigotimes_{\ell=1}^{n}{\!}_{q'}\,\ell\,,
\end{align}
where the leftmost side defines $n!_{q;\theta}$, while the rightmost side defines $n!_{q'}$.

The particular cases of $\theta=\pm 1$ were considered in \cite{Oikonomou07}, which led to two distinct definitions of the $q$-generalised factorial operations and, subsequently, to the two corresponding definitions of the $q$-multinomial coefficients.
These two cases are related by the first equation of (\ref{lam_symmetry_spec}), i.e.~the \q{additive duality} $q\leftrightarrow 2-q$.
More generally, by virtue of the $\theta$-symmetry, it holds that $n!_{q;\theta_{1}}=n!_{q';\theta_{2}}$ for arbitrary $\theta_{1},\,\theta_{2}\in\mathbb{R}\backslash\{0\}$, where $q'=f_{-\theta_{1}/\theta_{2}}(q)$.

\paragraph{The ${q}$-multinomial coefficient.}

Finally, the $q$-multinomial coefficient \cite{Suyari06,Oikonomou07}, which is the $q$-deformation of the multinomial coefficient (\ref{def:multinomial}), can be constructed based on the $q$-factorial (\ref{def:!q}), as well as the $q$-multiplication and $q$-division (\ref{def:q_multidiv}).
It is defined as follows:
\begin{equation}\label{def:q-multinomial1}
\binom{n}{\vK}_{\!q}\equiv\binom{n}{k_{1},\,\dots,\,k_{r}}_{\!q}\coloneqq 
n!_{q}\oslash_{q}\kko{\bigotimes_{i=1}^{r}{\!}_{q}\,k_{i}!_{q}}\,.
\end{equation}
It is easy to see that this reproduces the ordinary multinomial coefficient (\ref{def:multinomial}) in the limit $q\to 1$.
The definition given by Eq.~(\ref{def:q-multinomial1}) is explicitly expressed as
\begin{equation}\label{def:q-multinomial2}
\binom{n}{\vK}_{\!q}
=\bigg[h_{n}(q-1)-\sum_{i=1}^{r}h_{k_{i}}(q-1)+1\bigg]_{+}^{\f{1}{1-q}}
\end{equation}
in terms of the generalised harmonic number (\ref{def:harmonic}).

In the definition (\ref{def:q-multinomial1}), we used the $q$-factorial as defined in Eq.~(\ref{def:!q}).
If, instead, we used the $\theta$-generalised $q$-factorial (\ref{def:!q_gen}), we would obtain the same expression as in Eq.~(\ref{def:q-multinomial2}), with the only difference being that $q$ is replaced with $q'=f_{-\theta}(q)$, as defined in Eq.~(\ref{newlambda}).

\section{Emergent family of Tsallis entropies\label{sec:emergent}}

In this section, we derive our formula for $L_{q}(\P)$, defined continuously and smoothly across the entire real range of $q$. 
We will see that a family of Tsallis entropies emerges, characterised by an infinite sequence of entropic indices corresponding to the respective powers of $n$ in the formula's expansion coefficients.

\subsection{Derivation of the formula\label{subsec:derivation}}

Recall that the generalised harmonic number is related to the Riemann zeta function $\zeta(s)$ as its limit as $n\to\infty$:
\begin{equation}\label{def1:zeta}
\zeta(s)\coloneqq\lim_{n\to\infty}h_{n}(s)
=1+\f{1}{2^{s}}+\f{1}{3^{s}}+\dots\,.
\end{equation}
However, the Riemann zeta function defined in this way is well-defined only for $s\in\mathbb{C}$ such that $\mathfrak{R}(s)>1$.
Since we aim to explore the entire range of $q\equiv s+1\in\mathbb{R}$ in the context of the $q$-multinomial coefficient, we need a different approach.
This involves analytically continuing the Riemann zeta function to a meromorphic function across the entire complex plane, which exhibits a simple pole at $s=1$ with residue $1$.

For our purposes, the following alternative definition of the Riemann zeta function with $m\in\mathbb{N}$ is useful (see Appendix \ref{app:zeta_EM} for the derivation via the Euler–Maclaurin formula):
\begin{align}\label{def2:zeta}
\zeta(s)&\coloneqq 
h_{n}(s)-\f{n^{1-s}}{1-s}-\f{n^{-s}}{2}+{}\no\\
&\quad{}+\sum_{\ell=2}^{m}\binom{s+\ell-2}{\ell-1}\f{B_{\ell}}{\ell}n^{1-s-\ell}
-R_{n,m}(s)\,.
\end{align}
This expression applies for $s\in\mathbb{C}\backslash\{1\}$ and $\mathfrak{R}(s)>-m$, thus extending its domain by analytic continuation.
Here, $B_{\ell}$ denotes the $\ell$-th Bernoulli number (with $B_{1}=-\f{1}{2}$ and $B_{\ell}=0$ for odd $\ell\geq 3$), and $R_{n,m}(s)$ represents the remainder term:
\begin{equation}
R_{n,m}(s)=\binom{s+m}{m+1}\int_{n}^{\infty}\f{B_{m+1}(x-\floor*{x})}{x^{s+m+1}}\dd x\,,
\end{equation}
where $B_{\ell}(\cdot)$ denotes the $\ell$-th Bernoulli polynomial and $\floor*{x}$ represents the floor function that gives the largest integer less than or equal to $x$.
The remainder term $R_{n;m}$ vanishes in the limit $n\to\infty$, and for finite $n$, it vanishes for integer values of $s$ such that $-m\leq s\leq 0$.
Below, we only consider real values of $s=q-1\in\mathbb{R}$.
The second and third terms on the RHS of Eq.~(\ref{def2:zeta}) can be compactly expressed by combining them into a summation involving the Bernoulli numbers:
\begin{align}\label{def3:zeta}
h_{n}(q-1)&=\zeta(q-1)-\sum_{\ell=0}^{m}\binom{q+\ell-2}{\ell}\f{B_{\ell}}{q+\ell-2}\times{}\no\\
&\quad{}\times n^{2-q-\ell}+R_{n,m}(q-1)\,.
\end{align}
Eq.~(\ref{def3:zeta}) serves as the pivotal expression for deriving the formula for the asymptotic form of the $q$-multinomial coefficient.
By taking the $q$-logarithm of Eq.~(\ref{def:q-multinomial2}), we obtain
\begin{align}
\ln_{q}\binom{n}{\vK}_{\!q}
&=\f{1}{1-q}\bigg[h_{n}(q-1)-\sum_{i=1}^{r}h_{k_{i}}(q-1)\bigg]\label{lnq_multicoeff_almost1}\\
&\hspace{-2em}{}=\f{(r-1)\zeta(q-1)}{q-1}-\f{1}{1-q}\sum_{\ell=0}^{m}\binom{q+\ell-2}{\ell}\times{}\no\\
&\hspace{-2em}{}\times\f{B_{\ell}}{q+\ell-2}\bigg(n^{2-q-\ell}-\sum_{i=1}^{r}{k_{i}}^{2-q-\ell}\bigg)
+R_{m}(q-1)\,,\label{lnq_multicoeff_almost2}
\end{align}
where $R_{m}\coloneqq R_{n;m}-\sum_{i=1}^{r}R_{k_{i},m}$ collectively denotes the remainder terms.
By applying the definition of Tsallis entropy (\ref{def:Tsallis_ent}) and utilising the properties of the binomial coefficient, the second block of terms on the RHS of Eq.~(\ref{lnq_multicoeff_almost2}) can be reformulated as
\begin{align}\label{ceff}
&\sum_{\ell=0}^{m}C_{\ell}(q)n^{2-q-\ell}H_{2-q-\ell}(\{k_{i}/n\}_{i=1}^{r})\,,\no\\
&\qquad C_{\ell}(q)=\binom{q+\ell-1}{\ell}\f{B_{\ell}}{2-q-\ell}\,.
\end{align}
Therefore, in the asymptotic limit, Eq.~(\ref{lnq_multicoeff_almost2}) is expressed as, using $\P\equiv\vK/n$,
\begin{align}\label{formula_m}
L_{q,n;m}(\P)&\coloneqq
\f{\zeta(q-1)}{q-1}H_{0}(\P)+{}\no\\
&\hspace{-3em}{}+\sum_{\ell=0}^{m}C_{\ell}(q)n^{2-q-\ell}H_{2-q-\ell}(\P)+R_{m}(q-1)\,.
\end{align}
The first few coefficients $\{C_{\ell}(q)\}_{\ell\in\mathbb{Z}_{\geq 0}}$ are explicitly given by
\begin{align}\label{C_ell}
&C_{0}(q)=-\f{1}{q-2}\,,\quad 
C_{1}(q)=\f{q}{2(q-1)}\,,\no\\ 
&C_{2}(q)=-\f{q+1}{12}\,,\quad 
C_{4}(q)=\f{q(q+1)(q+3)}{720}\,,\dots
\end{align}
and $C_{\ell}(q)=0$ for odd integers $\ell\geq 3$ because the corresponding Bernoulli numbers vanish.
The asymptotic form (\ref{formula_m}) characterises Tsallis entropy within the $q$-deformed combinatorial framework.
For a single value of $q$ on the LHS, the RHS corresponds not to a single Tsallis entropy with a single entropic index, but rather to a family of Tsallis entropies as in Eq.~(\ref{formula_m}).
In other words, a series of measurements of the system's complexity (i.e.~the entropy), taken with \q{lenses} (i.e.~the entropy function, $H_{\circ}$) of various \q{magnifications} (i.e.~the set of entropic indices, $\{0\}\cup\{2-q-\ell\}_{\ell=0}^{m}$), contributes to the $q$-multinomial coefficient through the $q$-logarithm.

To ensure that the formula (\ref{formula_m}) is valid for all real values of $q$, we let $m\to\infty$ and denote $L_{q,n;m}$ in this limit as $L_{q}$ for notational simplicity.
The remainder terms also tend to zero in the asymptotic limit.
Consequently, the final expression of our formula takes the form of Eq.~(\ref{formula}) with the coefficients as defined in (\ref{ceff}).
Thus, the $q$-multinomial coefficient is associated with a family of Tsallis entropies characterised by a series of different degrees of nonadditivity, each scaling differently with respect to the system size ($n$).
As we shall see later, the first term, which is independent of $n$ and involves the zeta function, is crucial as it eliminates the divergences arising from $C_{0}(q)$ and $C_{1}(q)$ when $q=2$ and $q=1$, respectively.
From a combinatorial perspective, the natural emergence of Bernoulli numbers coupled with entropies in the formula based on the $q$-combinatorics is noteworthy, considering the limited discussion of Bernoulli numbers in relation to combinatorial approaches in the literature.

\subsection{Characteristic values and limits\label{subsec:behaviours}}

Let us explore the behaviour of $L_{q}$ across characteristic domains and values of $q$, with a particular emphasis on integer values.

When $q\in\mathbb{Z}_{\leq 0}$, $L_{q}$ can be expressed as a polynomial with a finite number of terms.
The highest power of $n$ is $n^{2-q}$, indicating that $L_{q}$ scales as $n^{-q}$ as $q$ tends towards negative infinity.
The least power of $n$ is $n^{2}$ when $q=-1$ and $q\leq 0$ is even, and $n^{3}$ when $q\leq -3$ is odd.
The number of nonvanishing terms is $1$ for $q=0$ and $2-\ceil*{\f{q}{2}}$ for $q\in\mathbb{Z}_{\leq -1}$, where $\ceil*{x}$ denotes the ceiling function, giving the smallest integer greater than or equal to $x$.
In particular, when $q=0$, $L_{q}$ simplifies to:
\begin{equation}\label{q->0}
{L}_{0}(\P)
=\f{n^{2}}{2}H_{2}(\P)\,,
\end{equation}
which is directly proportional to the second-order Tsallis entropy.\footnote{This result also directly follows from an elementary $q$-combinatorics calculation with $q = 0$. 
In Eq.~(\ref{lnq_multicoeff_almost1}), $h_{n}(-1)=\f{1}{2}n(n+1)$, and similarly for $k_{i}!_{q=0}$ for each $i$, where the linear terms cancel each other out.}
This case with $q=0$ on the LHS and $q'=2$ on the RHS corresponds to a particularly intriguing point from both information-theoretic and physical perspectives, especially when extending entropy to account for sensitivity to distances or affinities between the labelled categories $i=1,\,\dots,\,r$ \cite{Rao82,Leinster12,Okamura20}; see Section \ref{sec:affinity} for relevant discussion.

When $q\to 1$, the term involving the zeta function exhibits a pole at $q=1$ with residue $-\f{1}{2}(r-1)$, which is precisely cancelled by the $\ell=1$ component in the sum.
Consequently, ${L}_{q}$ remains well-defined and continuous at $q=1$, and is given by
\begin{align}\label{q->1}
{L}_{1}(\P)&=\lim_{q\to 1}{L}_{q}(\P)\no\\
&=nH_{1}+\f{1}{2}\ln\ko{\f{n}{\prod_{i=1}^{r}k_{i}}}-\f{1}{2}\ln(2\pi)H_{0}-{}\no\\
&\quad{}-\f{1}{6n}H_{-1}+\f{1}{90n^{3}}H_{-3}+\dots\,,
\end{align}
where $H_{q}\equiv H_q(\P)$ for simplicity.
This expression generalises the relation (\ref{ln_multicoeff}) seen in the Introduction.
The first three terms preceding the term involving $H_{-1}$ are straightforward consequences of the standard Stirling's formula, $\ln(n!)=\f{1}{2}\ln(2\pi)+\big(n+\f{1}{2}\big)\ln n-n+O(n^{-1})$, applied similarly to $\ln(k_{i}!)$.
The equation
\begin{equation}\label{pa_0}
\left.\f{\partial H_{q}(\P)}{\partial q}\right|_{q=0}=\sum_{i=1}^{r}\ln p_{i}+H_{0}(\P)
\end{equation}
is instrumental in deriving the second and third terms.
Interestingly, higher-order corrections, beyond those derivable from the standard Stirling's formula, are expressed by terms involving Tsallis entropy with negative odd values of $q$, coupled with corresponding powers of $n$. 

When $q\to 2$, the term involving the zeta function exhibits a pole at $q=2$ with residue $r-1$, which is precisely cancelled by the $\ell=0$ component in the sum, resulting in:
\begin{align}\label{q->2}
{L}_{2}(\P)&=\lim_{q\to 2}{L}_{q}(\P)\no\\
&=-\ln\ko{\f{n}{\prod_{i=1}^{r}k_{i}}}
+\gamma H_{0}+\f{1}{n}H_{-1}-{}\no\\
&\quad{}-\f{1}{4n^{2}}H_{-2}+\f{1}{24n^{4}}H_{-4}+\dots\,,
\end{align}
where $\gamma=\lim_{n\to\infty}(h_{n}(1)-\ln n)$ is Euler's constant.
Again, the relation (\ref{pa_0}) is useful in deriving the first and second terms.
Comparing Eq.~(\ref{q->2}) of our result with Eq.~(\ref{Sy_lnq_multi_b}) obtained in \cite{Suyari06}, it is evident that the latter amounts to extracting only the first term from the former, disregarding the constant term $\gamma H_{0}$ and all higher-order terms involving $H_{q\in\mathbb{Z}_{<0}}$.

Finally, when $q>2$, the term involving $H_{0}$ becomes the leading term in the expansion of $L_{q}$ with respect to $n$.
For sufficiently large $q\gg 2$, in the asymptotic limit, we have
\begin{equation}\label{q>>2}
{L}_{q\gg 2}(\P)\sim\f{\zeta(q-1)}{q-1}H_{0}\sim\f{1}{q}H_{0}\,.
\end{equation}
Subsequently, it is evident that $\lim_{q\to\infty}L_{q}(\P)=0$.

We also note when the function $L_{q}(\P)$ reaches its (relative) maximum in the asymptotic region (i.e.~$n\sim k_{i}\gg 1$ for all $i$).
First, observe that the term with the highest power in $n$ corresponds to the $\ell=0$ component in the sum, with its coefficient given by $C_{0}(q)H_{2-q}(\P)$.
When $q<2$, $C_{0}(q)>0$ and $H_{2-q}(\P)$ is concave; when $q>2$, $C_{0}(q)<0$ and $H_{2-q}(\P)$ is convex; and in the limit $q\to 2$, this term, together with the term involving the zeta function, gives the first two terms on the RHS of Eq.~(\ref{q->2}), whose leading term in $n$ is concave with respect to $\P$. 
These observations imply that $L_{q}(\P)$ is always concave with respect to $\P$, regardless of the value of $q$.
Therefore, the (relative) maximum value of $L_{q}(\P)$ in the asymptotic region is achieved in the equiprobable scenario, that is when $p_{i}=\f{1}{r}$ for all $i$.

\subsection{Comparison with results from the literature\label{subsec:comparison}}

\begin{figure}[!t]
\centering
\includegraphics[scale=0.62]{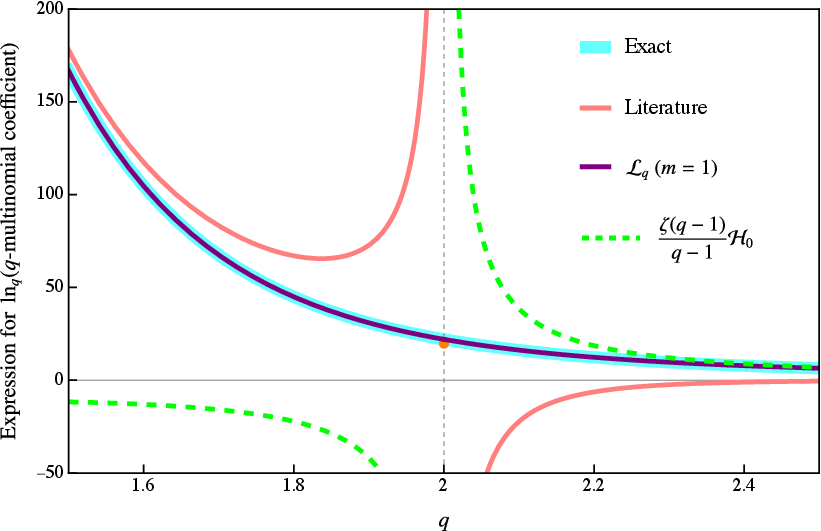}
\caption{{Comparison of expressions for the ${q}$-logarithm of the ${q}$-multinomial coefficient.}
The cyan line represents the exact expression (\ref{lnq_multicoeff_almost1}), while the purple line depicts our derived formula (\ref{formula}) with $m=1$.
For reference, the light green dashed line represents the first term in the formula, which involves the Riemann zeta function.
The pink line and the pink dot correspond to the expressions (\ref{Sy_lnq_multi}) as presented in the literature \cite[Eq.~(42)]{Suyari06}.
This visual comparison is based on the choices of $n=\sum_{i=1}^{5}k_{i}=1,866$, where $\vK=\{27,\,924,\,390,\,236,\,289\}$.}
\label{fig:qplot}
\end{figure}

Figure \ref{fig:qplot} visually compares these expressions for the $q$-logarithm of the $q$-multinomial coefficient.
Taking $r=5$ as an example, we generated five random numbers less than 1,000: $\vK=\{27,\,924,\,390,\,236,\,289\}$, resulting in $n=1,866$.
The cyan line represents the exact expression (\ref{lnq_multicoeff_almost1}) based on the definition (\ref{def:q-multinomial1}).
The pink line (for $q\neq 2$) represents the expression (\ref{Sy_lnq_multi_a}) from the literature \cite[Eq.~(42)]{Suyari06}, which does not accurately capture the behaviour of the $q$-logarithm of the $q$-multinomial coefficient near $q=2$. 
It diverges to $\pm\infty$ as $q$ approaches $2^{\pm 0}$ and yields negative values for $q>2$.
These discrepancies arise from the fact that the approximation used in the literature is essentially the following:
\begin{equation}\label{h_approx}
h_{n\gg 1}(q-1)~\stackrel{\raisebox{1ex}{\text{\footnotesize\cite{Suyari06}}}}{\approx}~
\int_{0}^{n}x^{1-q}\dd x=\f{n^{2-q}}{2-q}\,,
\end{equation}
which amounts to extracting only the term proportional to $n^{2-q}$ from the exact expansion.
This is equivalent to considering only the $\ell=0$ component from our formula (\ref{formula}), while ignoring the term involving the zeta function and any higher-order terms with $\ell\geq 1$.
The exact expansion is given by:
\begin{align}\label{h_exact}
h_{n}(q-1)&=\zeta(q-1)+\f{n^{2-q}}{2-q}+\f{n^{1-q}}{2}+{}\no\\
&\quad{}+\f{1-q}{12n^{q}}+\f{(q-1)q(q+1)}{720n^{2+q}}+\dots\,.
\end{align}
Although the approximation (\ref{h_approx}) becomes exact when $q=1$ (i.e.~$h_{n}(0)=\sum_{\ell=1}^{n}1=n$; notice that $\zeta(0)+\f{1}{2}=0$ in Eq.~(\ref{h_exact})), there is a discrepancy when we deviate from this value in either direction, whether large or small. 
Therefore, as long as the approximation (\ref{h_approx}) is used, the range of valid $q$ cannot encompass the entire set of real numbers.
This situation is essentially the same as in \cite[Eq.~(24)]{Oikonomou07}, where $q'=2-q$ is used instead of $q$ in Eq.~(\ref{h_approx}). 

In contrast, the purple line, representing our derived formula (\ref{formula}), aligns precisely with the exact result, including its behaviour around $q=2$.
This holds even when $m=1$, i.e.~when using only the first two terms corresponding to $\ell=0$ and $\ell=1$ in the sum.
For reference, the term involving the zeta function is represented by the light green dashed line. 
Visually, its graph confirms our findings in Section \ref{subsec:behaviours} that it plays a crucial role in neutralising the divergence arising from the expression in Eq.~(\ref{Sy_lnq_multi_a}), thereby yielding the exact formula.
Additionally, it can be seen that the term involving the zeta function becomes the primary contribution to the exact expression when the value of $q$ is sufficiently large ($q\gg 2$), as indicated in Eq.~(\ref{q>>2}).
Furthermore, as demonstrated earlier, our formula is inherently valid for $q<0$ by construction.
This reaffirms the potential utility and applicability of the formula across different contexts in information theory, thermostatistics or general physics, particularly when phenomena or systems associated with generalised multinomial coefficients near $q=2$ (or $q=0$, depending on convention) or $q<0$ are of interest.

\section{Extension to affinity-sensitive cases\label{sec:affinity}}

Based on the results obtained thus far, we offer some insights into potential avenues for extending our results.
Many studies have extended the entropy formula to account for the distance or similarity (affinity) between the constituent categories ($i=1,\,\dots,\,r$); see e.g.~\cite{Rao82,Leinster12,Okamura20} and references therein.
Such distance- or similarity-sensitive extended entropy includes Tsallis entropy (and BGS entropy, as its $q\to 1$ limit) as the limit of maximising distance or minimising affinity.
In the most general form, derived from an axiomatic approach, it is given by \cite[Eq.~(19)]{Okamura20}
\begin{widetext}
\begin{subnumcases}{\label{def:aff_entropy} 
\widetilde{H}_{q}(\P,\vA)\coloneqq}
\,\f{1}{1-q}\kko{\sum_{i=1}^{r}p_{i}\Bigg(\sum_{j=1}^{r}K(d_{ij})p_{j}\Bigg)^{q-1}-1} & \quad \text{$(q\neq 1)$}\,, \label{KO_a}  \\
\,-\sum_{i=1}^{r}p_{i}\ln\sum_{j=1}^{r}K(d_{ij})p_{j} & \quad \text{$(q=1)$}\,, \label{KO_b}
\end{subnumcases}
\end{widetext}
where $d_{ij}$ denotes the distance between two categories represented by $i$ and $j$ with convention $d_{ij}\in [0,1]$, with which the $r\times r$ distance matrix is given by $\vD\coloneqq [d_{ij}]_{i,j=1}^{r}$.
The kernel function $K(\cdot)$ is a continuous monotonically decreasing function satisfying $K(0)=1$ and $K(1)=0$.
The value $K(d_{ij})$ quantifies the affinity between two categories represented by $i$ and $j$, which defines the corresponding affinity matrix, $\vA\coloneqq [K(d_{ij})]_{i,j=1}^{r}$.
It is evident that $\lim_{q\to 1}\text{(\ref{KO_a})}=\text{(\ref{KO_b})}$.

In contrast to other formulations for distance- or similarity-based extended entropy in the literature \cite{Rao82,Leinster12}, the formula (\ref{def:aff_entropy}) provides two pathways to achieve the \q{zero-affinity limit}, where the affinity matrix simplifies to the identity matrix of size $r$, denotes as $\vI_{r}$.
This reduction causes the extended entropy (\ref{def:aff_entropy}) to converge either to Tsallis entropy ($q\neq 1$) or BGS entropy ($q=1$).
One pathway involves an \q{exogenous} reduction, where $d_{ij}\to 1-\delta_{ij}$ and $\delta_{ij}$ denotes the Kronecker delta.
The other pathway is an \q{endogenous} reduction, such as taking the limit $\alpha\to 0$ of the kernel function $K(d_{ij})=1-(d_{ij})^{\alpha}$ while keeping $d_{ij}$ unreduced.

Now, recall the definition of the ordinary multinomial coefficients (\ref{def:multinomial}), the case with $q=1$.
This quantity can be directly interpreted in combinatorial terms as the number of ways to distribute $n$ distinct objects into $r$ distinct categories, with $k_{i}$ objects in the $i$-th category ($i=1,\,\dots,\,n$).
Here, the concept of \q{distinct categories} corresponds to an affinity matrix equalling the identity matrix. 
However, in the case of a general affinity matrix, where there is an arbitrary affinity between categories with different labels ($i\neq j$), it remains to be clarified how the notion of a multinomial coefficient can be generalised in a mathematically consistent and physically relevant manner.
In the physical context mentioned in the Introduction, this problem involves quantifying the \emph{effective} number of possible microstates associated with a given macrostate, considering \q{correlations} among $r$ states where $n$ particles are distributed.

We argue that a solution to this problem involves substituting $H_{q}(\P)$ in the formula (\ref{formula}) with $\widetilde{H}_{q}(\P,\vA)$ given in Eqs.~(\ref{def:aff_entropy}) as follows:
\begin{align}\label{formula_aff}
\ln_{q}\widetilde{\binom{n}{\vK|\vA}}_{\!q}
&\equiv\widetilde{L}_{q}(\P,\vA)
\coloneqq\f{\zeta(q-1)}{q-1}\widetilde{H}_{0}(\P,\vA)+{}\no\\
&\quad{}+\sum_{\ell=0}^{\infty}C_{\ell}(q)n^{2-q-\ell}\widetilde{H}_{2-q-\ell}(\P,\vA)\,,
\end{align}
where quantities with a tilde symbol indicate their dependence on distance or affinity.\footnote{Note that the affinity-based extended entropy with $q=0$, given by $\widetilde{H}_{0}(\P,\vA)=\sum_{i=1}^{r}p_{i}\big/\sum_{j=1}^{r}p_{j}K(d_{ij})-1$, depends on the specific distribution $\P$, in contrast to its zero-affinity case that depends only on $\abs{\P}$.}
In the zero-affinity limit, where $\vA\to \vI_{r}$ or $K(d_{ij})\to\delta_{ij}$ for all $i,\,j$, this formula simplifies to the one presented in Eq.~(\ref{formula}), giving $\ln_{q}\widetilde{\binom{n}{\vK|\vI_{r}}}_{\!q}\equiv\ln_{q}\binom{n}{\vK}_{\!q}=L_{q}(\P)$.
Conversely, when $K(d_{ij})\to 1$ for all $i,\,j$, the formula (\ref{formula_aff}) evaluates to $0$.
This indicates that the corresponding affinity-based extended $q$-multinomial coefficient is ($\exp_{q}0=$) $1$, representing the scenario with a single category.

Consider a more general case that encompasses these two scenarios.
For instance, take an affinity matrix formed by the direct sum of an identity matrix of size $a\,(\leq r)$, $\vI_{a}$, and a matrix of size $r-a$, $\vB_{r-a}$, where all elements equal to $1$.
The formula (\ref{formula_aff}) is then calculated as follows:
\begin{equation}\label{I+B}
\widetilde{L}_{q}(\P,\vI_{a}\oplus\vB_{r-a})
=(P_{a})^{q}L_{q}(\P_{a})+L_{q}(\{P_{a},1-P_{a}\})\,,
\end{equation}
where $\P_{a}\coloneqq\{p_{i}/P_{a}\}_{i=1}^{a}$ with $P_{a}\coloneqq\sum_{i=1}^{a}p_{i}$.
The case where $a=r$ corresponds to the previously discussed zero-affinity case, while where $a=0$ corresponds to the zero-entropy (single-category) case.
If we relabel and set $p_{a+1}\equiv 1-P_{a}$, this case represents a system with $a+1$ maximally distinct categories with probabilities $\P'\equiv \{p_{1},\,\dots,\,p_{a+1}\}$.
It is easy to check that Eq.~(\ref{I+B}) equals $L_{q}(\P')$, as it should be.

Furthermore, a particularly interesting case arises when $q=0$ for the quantity (\ref{formula_aff}).
In this case, we obtain
\begin{align}
\widetilde{L}_{0}(\P,\vA)
&=\f{n^{2}}{2}\widetilde{H}_{2}(\P,\vA)\no\\
&=\f{n^2}{2}\bigg(1-\sum_{i=1}^{r}\sum_{j=1}^{r}p_{i}p_{j}K(d_{ij})\bigg)\,,
\end{align}
extending the relation (\ref{q->0}) to the affinity-sensitive case.
It represents essentially the same quantity first considered in \cite{Rao82} by setting the kernel function as $K(d_{ij})=1-d_{ij}$.
Notably, this combinatorial quantity adheres to the Nesting Principle \cite[Section 2.3; Fig.~3]{Okamura20}, which asserts that it remains invariant under any arbitrary (re-)grouping of the constituent categories, provided a newly defined proper set of distances is applied \cite[Eq.~(30)]{Okamura20}.
To understand this mechanism, consider grouping the $r$ categories we have discussed into fewer $r'$ categories by internally grouping them (coarse-graining).
Specifically, for a set of integers $\{I_{i'}\}_{i'=0}^{r'}$ that satisfy $I_{0}\equiv 1<I_{1}\leq\dots\leq I_{r'-1}<r\equiv I_{r'}$, define $S_{i'}\coloneqq\{i\in\mathbb{N}\,|\,I_{i'-1}<i\leq I_{i'}\}$ for $i'=1,\,\dots,\,r'$.
A new set of integers at the aggregated level is then defined by $\vK_{\#}\coloneqq\{k'_{1},\,\dots,\,k'_{r'}\}$, where $k'_{i'}=\sum_{i\in S_{i'}}k_{i}$ and $\sum_{i'=1}^{r'}k'_{i'}=\sum_{i=1}^{r} k_{i}=n$.
The corresponding probability distribution is defined by $\P_{\#}\coloneqq\vK_{\#}/n$.
Additionally, at this aggregated level, the new distance matrix is defined as $\vD_{\#}=[d_{i'j'}]_{i',j'=1}^{r'}$, where $d_{i'j'}\coloneqq\sum_{i\in S_{i'}}\sum_{j\in S_{j'}}\f{k_{i}k_{j}}{k'_{i'}k'_{j'}}d_{ij}$, and the new affinity matrix is defined as $\vA_{\#}=[K(d_{i'j'})]_{i',j'=1}^{r'}$, employing the specific kernel function $K(x)=1-x^{\alpha}$, where $\alpha>0$.
Then, the Nesting Principle for the $(q=)\,0$-generalised multinomial coefficient is manifested by the following identity:
\begin{align}\label{nesting_inv}
&\widetilde{L}_{0}(\P,\vA)
\equiv\widetilde{L}_{0}(\P_{\#},\vA_{\#})\,,\no\\
&~~\text{or}~~
\widetilde{\binom{n}{\vK|\vA}}_{0}\equiv\widetilde{\binom{n}{\vK_{\#}|\vA_{\#}}}_{0}\,.
\end{align}

\section{Summary and conclusion\label{sec:conclusion}}

In this paper, we revisited the problem of deriving a formula for $q$-generalised multinomial coefficients within the $q$-deformed algebraic framework \cite{Nivanen03,Borges04}, which is central in nonextensive statistical mechanics. 
Our derived formula (\ref{formula}), in contrast to the expressions found in the existing literature \cite{Suyari06,Oikonomou07}, offers an exact, smooth function valid for all real values of $q$.
An essential technique for achieving this derivation is the analytic continuation of the Riemann zeta function, which stems from the $q$-deformed factorials.
The resulting formula is represented as an infinite series expansion with respect to the system size $n$, combined with Tsallis entropies, where the power of $n$ and the indices of the Tsallis entropies match.
The expansion coefficients involve Bernoulli numbers, which is intriguing from a combinatorial perspective.
Thus, our formula provides a distinctive characterisation of Tsallis entropy within the $q$-deformed combinatorial framework.

We also underscored the special invariance property inherent in the $q$-algebra induced by the transformation (\ref{newlambda}).
Various arbitrary values of the entropic parameter, which lead to seemingly distinct formulations of $q$-deformed quantities such as the $q$-factorial and the $q$-multinomial coefficient, are shown to be related by this symmetry, indicating they represent the same underlying theory.

Furthermore, we extended our results to encompass more general, distance- or affinity-sensitive cases along the lines of \cite{Rao82,Leinster12,Okamura20}.
This addresses the proposal to extend the definition of the multinomial coefficient to accommodate arbitrary distances or affinities between different categories, ensuring mathematical consistency and physical relevance.
In this context, we also highlighted that the multinomial coefficient deformed by the $(q=)\,0$-algebra satisfies a special grouping invariance known as the Nesting Principle \cite{Okamura20}.

\begin{acknowledgments}
I would like to thank an anonymous referee of \textit{Physics Letters A} and Professor Hiroki Suyari for their comments on the manuscript.
The views and conclusions contained herein are those of the author and should not be interpreted as necessarily representing the official policies or endorsements, either expressed or implied, of any of the organisations with which the author is currently or has been affiliated in the past.
\end{acknowledgments}

\subsection*{\textsc{NOTES}}
The published version of this article, \textit{Phys.~Lett.~A} \textbf{525}, 129912 (2024), is available at \url{https://doi.org/10.1016/j.physleta.2024.129912}.
The Graphical Abstract for this study can also be accessed in the Zenodo repository at \url{https://doi.org/10.5281/zenodo.13894321}.

\vspace{1eM}
                  \appendix                %

\section{More on the \textit{q}-algebra\label{app:q-algebra}}

\paragraph{The ${q}$-addition and ${q}$-subtraction.}

In addition to $q$-multiplication and $q$-division discussed in the main text, $q$-addition and $q$-subtraction \cite{Nivanen03,Borges04} also form an essential foundation for nonextensive statistics \cite{Tsallis88,Curado91,Abe01}. 
These operations are defined between two real numbers $x$ and $y$ with $q\in\mathbb{R}\backslash\{1\}$ as follows:
\begin{align}
x&\oplus_{q}y\coloneqq 
x+y+(1-q)xy\,,\label{def:q_plus}\\
x&\ominus_{q}y\coloneqq 
\f{x-y}{1+(1-q)y}\qquad 
(1+(1-q)y\neq 0)\,,\label{def:q_minus}
\end{align}
which become ordinary addition and subtraction in the limit $q\to 1$, respectively.
Although found in no prior literature, the $q$-addition and $q$-subtraction defined above can equivalently be defined as follows:
\begin{equation}\label{def:q_addsubt}
x\,\Big\{{}^{\displaystyle\oplus_{q}}_{\displaystyle\ominus_{q}}\Big\}\,y
\coloneqq\f{[1+(1-q)x][1+(1-q)y]^{\pm 1}-1}{1-q}\,,
\end{equation}
where the upper signs on the RHS correspond to the definition of $q$-addition ($\oplus_{q}$) and the lower signs correspond to that of $q$-subtraction ($\ominus_{q}$).
The $q$-addition is commutative ($x\oplus_{q}y=y\oplus_{q}x$), associative ($x\oplus_{q}(y\oplus_{q}z)=(x\oplus_{q}y)\oplus_{q}z$), but not distributive with respect to ordinary multiplication ($\alpha(x\oplus_{q}y)\neq \alpha x\oplus_{q}\alpha y$).
It also follows that
\begin{align}\label{q_sum_n}
&\bigoplus_{i=1}^{n}{\!}_{q}\,x_{i}
=\ln_{q}\prod_{i=1}^{n}\exp_{q}x_{i}
=\f{\prod_{i=1}^{n}[1+(1-q)x_{i}]-1}{1-q}\no\\
&\quad\Rightarrow\quad 
\exp_{q}\bigoplus_{i=1}^{n}{\!}_{q}\,x_{i}
=\prod_{i=1}^{n}\exp_{q}x_{i}\,.
\end{align}

\paragraph{The symmetry.}

As with $q$-multiplication and $q$-division, there exists a symmetry \cite{Korbel17} regarding $q$-addition and $q$-subtraction:
\begin{equation}\label{lam_symmetry2}
\f{(\theta x)\odot_{q}(\theta y)}{\theta}
=x\odot_{q'}y~\quad 
(\odot=\oplus,\,\ominus)
\end{equation}
with the same transformation of the entropic parameter as in Eq.~(\ref{newlambda}), defined with $\theta\in\mathbb{R}\backslash\{0\}$.
The relation (\ref{lam_symmetry2}) can be viewed as a $q$-generalisation of the distributive law, which in the case of $q=1$ takes the form $\theta x\pm\theta y=\theta (x\pm y)$.
We also point out that $q$-addition and $q$-subtraction can be defined equivalently as binary operations that satisfy the following functional equations:
\begin{equation}\label{def2:q_addsubt}
f_{x}(q)f_{y}(q)=f_{x\oplus_{q}y}(q)\,,\qquad 
\f{f_{x}(q)}{f_{y}(q)}=f_{x\ominus_{q}y}(q)\,,
\end{equation}
where $f_{\circ}$ is the same transformation as defined in Eq.~(\ref{newlambda}).

\paragraph{Relations involving the ${q}$-logarithmic/exponential functions.}

Some relations involving the $q$-logarithmic and $q$-exponential functions between $q$-addition, $q$-subtraction, $q$-multiplication and $q$-division are given:
\begin{subequations}
\begin{align}
\ln_{q}(xy)&=\ln_{q}x\oplus_{q}\ln_{q}y\no\\
	&\hspace{-3em}{}=\ln_{q}x+\ln_{q}y+(1-q)\ln_{q}x\ln_{q}y\,,\label{lnq_x*y}\\
\ln_{q}(x/y)&=\ln_{q}x\ominus_{q}\ln_{q}y\no\\
	&\hspace{-3em}{}=\big(\ln_{q}x-\ln_{q}y\big)\big/\big[1+(1-q)\ln_{q}y\big]\,,\label{lnq_x/y}\\
\ln_{q}(x\otimes_{q}y)&=\ln_{q}x+\ln_{q}y\,,\label{lnq_x*qy}\\
\ln_{q}(x\oslash_{q}y)&=\ln_{q}x-\ln_{q}y\,,\label{lnq_x/qy}\\
\exp_{q}x\cdot\exp_{q}y&=\exp_{q}(x\oplus_{q}y)\no\\
	&\hspace{-3em}{}=\exp_{q}\big[x+y+(1-q)xy\big]\,,\label{expq_x*y}\\
\exp_{q}x\big/\exp_{q}y&=\exp_{q}(x\ominus_{q}y)\no\\
	&\hspace{-3em}{}=\exp_{q}\big[(x-y)\big/\big(1+(1-q)y\big)\big]\,,\label{expq_x/y}\\
\exp_{q}x\otimes_{q}\exp_{q}y&=\exp_{q}(x+y)\,,\label{expq_x*qy}\\
\exp_{q}x\oslash_{q}\exp_{q}y&=\exp_{q}(x-y)\,.\label{expq_x/qy}
\end{align}
\end{subequations}
Other properties of the $q$-algebra and the $q$-calculus, including the $q$-derivative and the $q$-integral, are discussed in literature \cite{Nivanen03,Borges04,Borges22}.

\section{Derivation of Eq.~(\ref{def2:zeta})\label{app:zeta_EM}}

In deriving Eq.~(\ref{def2:zeta}) in the main text, the key lies in the Euler-Maclaurin formula presented below:
\begin{theorem}
Let $a$ and $b$ be integers such that $a<b$, and let $f:[a,b]\to\mathbb{R}$ be continuous.
Then, for all integers $m\in\mathbb{Z}_{\geq 0}$, if $f$ is a $C^{m+1}$ function,
\begin{align}\label{EM}
\sum_{j=a}^{b}f(j)&=\int_{a}^{b}f(x)\dd x
+\f{1}{2}\ko{f(a)+f(b)}+{}\no\\
&\hspace{-2em}{}+\sum_{k=1}^{m}\f{(-1)^{k+1}B_{k+1}}{(k+1)!}\big[f^{(k)}(b)-f^{(k)}(a)\big]+{}\no\\
&\hspace{-2em}{}+\f{(-1)^{m}}{(m+1)!}\int_{a}^{b}B_{m+1}(x-\floor*{x})f^{(m+1)}(x)\dd x\,,
\end{align}
where $B_{\ell}$ is the $\ell$-th Bernoulli number with convention $B_{1}=-\f{1}{2}$, $B_{\ell}(x)$ is the $\ell$-th Bernoulli polynomial of $x$ and $\floor*{x}$ is the floor function which gives the largest integer less than or equal to $x$.
\end{theorem}
We apply this formula to the function $f(x)=x^{-s}$, $\mathfrak{R}(s)>1$, with $a=n\in\mathbb{N}$ and $b\to\infty$.
Noting that 
\begin{align}
f^{(k)}(x)&=(-s)(-s-1)\cdots(-s-k+1)x^{-s-k}\no\\
&=\f{(s+k-1)!}{(s-1)!}\f{(-1)^{k}}{x^{s+k}}\,,\no
\end{align}
the RHS of Eq.~(\ref{EM}) transforms into:
\begin{align}\label{EM:RHS}
&\int_{n}^{\infty}\f{dx}{x^{s}}
+\f{n^{-s}}{2}
-\sum_{k=1}^{m}\f{(-1)^{k+1}B_{k+1}}{(k+1)!}\cdot\f{(s+k-1)!}{(s-1)!}\f{(-1)^{k}}{n^{s+k}}+{}\no\\
&{}\quad+\f{(-1)^{m}}{(m+1)!}\int_{n}^{\infty}B_{m+1}(x-\floor*{x})\f{(s+m)!}{(s-1)!}\f{(-1)^{m+1}}{x^{s+m+1}}\dd x\no\\
&{}=-\f{n^{1-s}}{1-s}
+\f{n^{-s}}{2}
+\sum_{k=1}^{m}\binom{s+k-1}{k}\f{B_{k+1}}{k+1}n^{-s-k}-{}\no\\
&{}\quad -\binom{s+m}{m+1}\int_{n}^{\infty}\f{B_{m+1}(x-\floor*{x})}{x^{s+m+1}}\dd x\,.
\end{align}
On the other hand, the LHS of Eq.~(\ref{EM}) becomes
\begin{equation}\label{EM:LHS}
\f{1}{n^{s}}+\f{1}{(n+1)^{s}}+\f{1}{(n+2)^{s}}+\dots
=\zeta(s)-h_{n}(s)+\f{1}{n^{s}}\,.
\end{equation}
Subsequently, Eq.~(\ref{def2:zeta}) follows from equating Eqs.~(\ref{EM:RHS}) and (\ref{EM:LHS}).
The resulting expression for $\zeta(s)$ applies for $s\in\mathbb{C}\backslash\{1\}$ and $\mathfrak{R}(s)>-m$, thus extending its domain by analytic continuation.
By taking the limit as $m\to\infty$, it can be further analytically continued to $\mathbb{C}\backslash\{1\}$, where $\zeta(s)$ has a simple pole at $s=1$.
This formulation of the Riemann zeta function with $n=1$ is often utilised to derive various well-known summation results: $\zeta(0)=1+1+1+\dots=-\f{1}{2}$, $\zeta(-1)=1+2+3+\dots=-\f{1}{12}$, and more generally, $\zeta(-m)=-\f{B_{m+1}}{m+1}$ for $m\in\mathbb{Z}_{\leq 0}$.

\bibliographystyle{apsrev4-2}

%

\end{document}